\documentclass[12pt]{article}

\usepackage{sectsty}
\usepackage{graphicx}
\usepackage{cite}
\usepackage{url}
\usepackage{lscape}
\usepackage{indentfirst}
\usepackage{latexsym}
\usepackage{multirow}
\usepackage{tabls}
\usepackage{wrapfig}
\usepackage{longtable}
\usepackage{supertabular}
\usepackage{caption}
\usepackage{subcaption}
\usepackage{amsmath}
\usepackage{amssymb}
\usepackage{setspace}
\usepackage[bookmarks, colorlinks=true, plainpages=false, citecolor=blue, urlcolor=blue, filecolor=blue, linkcolor=blue]{hyperref} 
\usepackage[capitalise,nameinlink]{cleveref}
\usepackage[font={footnotesize,stretch=1.0}]{caption}

\allsectionsfont{\centering}
\newcommand{\pd}[2]{\frac{\partial #1}{\partial #2}}

\usepackage{multicol}
\usepackage{authblk}
\usepackage[margin=1.0in]{geometry}

\title{Demonstration of OpenMC as a framework for atomic transport and plasma interaction}

\author[1]{G. J. Wilkie}
\author[2]{P. K. Romano}
\author[1]{R. M. Churchill}
\affil[1]{Princeton Plasma Physics Laboratory, Princeton, NJ, USA}
\affil[2]{Argonne National Laboratory, Lemont, IL, USA}

\begin{document}

\maketitle

\begin{abstract}
   Modern tooling is demanded for predicting the transport and reaction characteristics of atoms and molecules, especially in the context of magnetic confinement fusion. DEGAS2, among the most common and capable tools currently in use, shares many fundamental similarities with the OpenMC framework, which was primarily developed for neutron and photon transport. In this work, we demonstrate that OpenMC is suitable for atomic transport calculations. The relative error between the models is small, and the performance of OpenMC is at least comparable to DEGAS2. This is the case even without taking advantage of heterogeneous computing architecture, which is only one of the several remarkable new capabilities that this demonstration heralds.
\end{abstract}

\section{Background and motivation}

The Monte Carlo method has been recognized as a robust and efficient method for estimating transport and reaction properties of particles. It has been a mainstay technique in neutron transport\cite{spanier_monte_2008}, especially in the context of nuclear fission. As research in magnetic confinement for nuclear \emph{fusion} progressed, the importance of atomic and molecular (``neutral'') transport and reactions was recognized. Predicting the evolution of a magnetically confined plasma is governed by a set of coupled partial differential equations, and plasma interaction with neutrals account for source and sink terms therein. Therefore, robustness of the neutral model is a particularly important asset, since it is run repeatedly with little interaction from the user, who is likely most interested in the evolution of the plasma. 

Tools for predicting the evolution of neutrals in magnetic confinement fusion were developed over the past half century, borrowing techniques from neutron transport simulation. Currently, the state-of-the-art is represented by two Monte Carlo solvers: EIRENE\cite{reiter2005eirene} and DEGAS2\cite{stotler_neutral_1994}. The former is closely associated with the SOLPS family of software, and the latter has a history of coupling to UEDGE \cite{joseph_coupling_2017} and XGC \cite{wilkie2024reconstruction} (fluid and kinetic edge plasma solvers, respectively). In addition to robust coupling to plasma solvers, neutral simulations are useful in their own right by interpreting experimental diagnostics or performing first-principles studies of atomic physics and plasma interaction. 

As capable as these legacy Monte Carlo tools are, the fusion community could benefit from an open source simulation framework for neutrals that: utilizes modern programing languages; easily parallelizes on modern CPU/GPU computational architectures; and integrates well with other software libraries.
Indeed, several candidate frameworks and algorithms have recently been proposed \cite{mijin2021sol,bernard_kinetic_2022,gonzalez2022comparison,wilkie_multidisciplinary_2023}. However, none of these are currently capable of matching the generality and/or robustness of EIRENE and DEGAS2.

The neutronics community has evolved to modern software standards. Developed under the Exascale Computing Project, OpenMC~\cite{romano2015ane1} has emerged as the eminent open source software for Monte Carlo neutron transport calculations. Unsurprisingly, the fundamental algorithm in OpenMC shares many similarities with the neutral transport solvers elaborated upon above. However, it comes with several modern improvements including: powerful application programming interfaces (APIs), GPU acceleration, and integration of CAD-based geometry. 

This work explores the potential for adapting OpenMC for atomic transport calculations, comparing the performance and accuracy to DEGAS2. Doing so for these test cases requires only modest changes to OpenMC, performed on a separate fork. Importing the full species, reaction, and source generality of DEGAS2 is beyond the scope of the present work. The goal here is to demonstrate a proof-of-principle that the structure of OpenMC is consistent with the needs of neutral transport in a fusion context, both in terms of performance and accuracy.

\Cref{sec:formulation} provides a brief overview of the physical problem being solved: steady-state solution of the linear Boltzmann equation. A review of the specific computational tools OpenMC and DEGAS2 follows in \cref{sec:methods}, along with descriptions of the common geometric representation and treatment of atomic reactions. \Cref{sec:results} contains the main results, showing benchmarks for simple and complicated spatial domains considering different combinations of reactions. Finally, our vision for the path forward is described in \cref{sec:future}.

\section{The Boltzmann equation for neutron and atomic tranport} \label{sec:formulation}

This section outlines the fundamental physics to be solved: the Boltzmann transport equation, common to both neutron and atomic/molecular transport. 
%, is recast as an integral equation to make it amenable to solution by the Monte Carlo integration method. The formal exposition borrows from Ref. \cite{spanier_monte_2008} and the reader is directed there for further details. 
We wish to find the steady-state phase space distribution for species $s$: $f_s\left( \mathbf{r}, \mathbf{v} \right)$. The distribution function is defined to be the probability of finding a particle in an infinitesimal volume spanned by $\mathbf{r}$ and $\mathbf{r} + \mathrm{d}^3\mathbf{r}$ with a velocity within $\mathbf{v}$ and $\mathbf{v} + \mathrm{d}^3\mathbf{v}$. By convention, it is normalized such that the particle density is $n_s\left(\mathbf{r} \right) = \int f_s \mathrm{d}^3\mathbf{v}$.  Under the assumptions of instantaneous binary collisions and molecular chaos, the distribution is found by solving the linearized Boltzmann equation:
\begin{equation} \label{boltzmanneq}
   \mathbf{v} \cdot \nabla f_s = S(\mathbf{r}, \mathbf{v}) - L(\mathbf{r}, \mathbf{v}) f_s + \sum\limits_{s' \neq s} C\left[ f_s, f_{s'} \right],
\end{equation}
where terms representing the sources and sinks of particles are given by $S$ and $L$, respectively. The \emph{collision operator} is $C$ and can be defined in terms of the differential scattering cross section $\frac{\partial \sigma_{ss'}}{\partial \Omega}$ as:
\begin{equation}
   C\left[ f_s, f_{s'} \right] \equiv \int u \pd{\sigma_{ss'}}{\Omega} \left[ f_s(\mathbf{v}') f_{s'}(\mathbf{w}') - f_s(\mathbf{v}) f_{s'}(\mathbf{w}\right) \mathrm{d}^2 \boldsymbol{\Omega} \mathrm{d}^3 \mathbf{w}.
\end{equation}
The collision operator remains a function of $\mathbf{v}$; the target velocities $\mathbf{w}$ have been intergrated over. The relative speed between collision partners (invariant for elastic collisions) is $u\equiv |\mathbf{v}-\mathbf{w}|$. The solid angle $\boldsymbol{\Omega}$ is a function of scattering angle, which also determines the post-collision velocities $\mathbf{v}'$ and $\mathbf{w}'$ in terms of the pre-collision velocities $\mathbf{v}$ and $\mathbf{w}$. For the special case of charge exchange (CX) between atoms (species $s$) and like-nucleus ions (species $s'$), then $\mathbf{w}' = \mathbf{v}$ and $\mathbf{v}' = \mathbf{w}$. The differential cross section can then be rigorously integrated to obtain:
\begin{equation}
   C_{CX}\left[ f_s, f_{s'} \right] \equiv \int u \sigma_{ss'}(u) \left[ f_s(\mathbf{w}) f_{s'}(\mathbf{v}) - f_s(\mathbf{v}) f_{s'}(\mathbf{w}\right) \mathrm{d}^3 \mathbf{w}.
\end{equation}
with the total cross section $\sigma_{ss'}$. The inclusion of atom-ion elastic scattering, which is quantum-mechanically indistinguishable from charge-exchange \cite{krstic_consistent_1999} complicates this picture but can readily be included even in the scattering-angle integrated approach by choosing an appropriate angular weighting for the elastic scattering part (e.g., momentum exchange). 

Stochastic methods are an attractive choice to solve \cref{boltzmanneq} to mitigate the high dimensionality of the problem and to enable the rigorous treatment of curved surfaces (see \cref{geomSec}). Deterministic methods have also been fruitfully explored \cite{pitchford_comparative_1982,hagelaar_solving_2005,gamba_galerkinpetrov_2018,wilkie_multidisciplinary_2023} and are especially attractive in the transition regime of moderate collisionality. In the highly collisional regime, the fluid approximation is most appropriate \cite{chapman_mathematical_1990}. However, stochastic methods remain robust, rigorous, and efficient in the free-streaming regime, where scattering events are rare compared to absorption.  

In order to take advantage of Monte Carlo integration in solving \cref{boltzmanneq}, it must be transformed into a strict integral equation by applying an integrating factor over phase space (filtered by some function $M(\mathbf{r},\mathbf{v})$). The choice of this integration kernel is known as a \emph{tally}. Typically, the spatial dependence of $M$ is that of a simple filter: unity in some region and zero elsewhere. 
%When subdividing the computational domain, each finite volume becomes its own filter and particle trajectories (\emph{flights}) are shared to calculate all relevant tallies. 
The velocity dependence of $M$ depends on which moment of the distribution function is of interest to the user. Unity would give the local particle density, $m_s \mathbf{v}$ the momentum density (where $m_s$ is the species mass), $\tfrac{1}{3} m_s v^2$ the pressure, etc. In this way, the particular form that \cref{boltzmanneq} takes the form of a Fredholm integral equation of the second kind for the tally moment. See Refs. \cite{spanier_monte_2008,reiter2005eirene} for details.

\section{Overview of methods} \label{sec:methods}

\subsection{OpenMC}

OpenMC is a community-developed, open source framework for Monte Carlo particle transport simulation that has been developed primarily for applications in nuclear energy (fission and fusion). It supports particle transport on both a native constructive solid geometry representation as well as CAD-based geometries through the DAGMC library \cite{wilson2010acceleration}. For neutron and photon transport, it relies on either continuous-energy cross section data or multigroup cross sections. Parallelism is enabled via a hybrid MPI and OpenMP programming model. The core transport solver in OpenMC is written in C++ while a Python API enables sophisticated workflows, including programmatic generation of input, analysis of results, visualization, and interfaces to underlying data files. OpenMC also features a built-in depletion/activation solver~\cite{romano2021ane} that can predict the change in material compositions over time due to neutron irradiation.

OpenMC has seen significant growth in its feature set as well as its user/developer community over the last 10 years. In particular, many features have been added to support applications in fusion energy, including the ability to model torii, support for CAD-based geometry, fixed-source activation calculations~\cite{romano2020anl}, and shutdown dose rate calculations~\cite{peterson2024nf}. Consequently, it has become quickly adopted by a significant fraction of the nascent private fusion industry. As of 2024, the codebase has well over 100 contributors.

In addition to feature expansion, many of the R\&D activities around OpenMC have focused on performance and scalability. Under the U.S. Exascale Computing Project, a branch of the code was ported to execute on GPUs using the target offload model from the OpenMP 4.5+ standard~\cite{tramm_sna_mc_2024}. The GPU port has been utilized on two of the largest supercomputers in the world, the Aurora machine at Argonne National Laboratory and the Frontier machine at Oak Ridge National Laboratory, reaching a throughput of over 1 billion particles per second on a challenging nuclear reactor problem. Other optimizations~\cite{romano2025cad} have led to significant performance improvements on complex models such as the E-lite model~\cite{juarez2021nature} of the ITER experiment.

\subsection{DEGAS2}

DEGAS2 is one of the two premier global tools for estimating plasma-neutral interaction in the context of magnetic confinement fusion \cite{stotler_neutral_1994}. It was written as an upgrade of the original DEGAS code and shares many similarities, particularly its fundamental algorithm, with its counterpart EIRENE and, indeed, OpenMC.

\Cref{boltzmanneq} is generalized in two ways in DEGAS2 and EIRENE to account for transient evolution and nonlinear scattering. Transience is accounted for by choosing a time integration method allowing the previous timestep's solution to act as a source. For example, forward Euler time integration over a step $\Delta t$ to find the tallies at step $n$ yields additional terms $f^{n-1}/\Delta t$ and $1/\Delta t$ in $S$ and $L$, respectively. Nonlinear scattering is approximated iteratively and with a simple Bhatnagar--Gross--Krook (BGK) collision operator against an ``effective Maxwellian'' from the previous iteration \cite{kotov_numerical_2007}. 

One of the features of DEGAS2 that distinguishes it from other codes like OpenMC is the generality with respect to species and reactions. It includes an extendable database of many reactions and wall-interaction models for hydrogenic atoms, molecules, and other impurities. Rates of reactions with electrons are pre-calculated by integrating over the cross section as a tabulated function of electron density and temperature, assuming electron velocities are well-separated from the atoms. The effect of electron-impact excitation is accounted for in a collisional radiative model (CRM), which assumes excited states are short-lived compared to their transport (either from spontaneous de-excitation or by frequent electron collisions). As such, the excitation reactions are part of the ionization rates and associated statistics (e.g., energy lost per ionization).

For the sake of comparison to OpenMC, we are limiting the test particles species to be hydrogen undergoing electron-impact ionization/excitation and charge exchange with protons. Generalizing OpenMC with respect to species to reproduce the capability of DEGAS2 will require significant effort that we hope is motivated by this work.

\subsection{Geometrical representation} \label{geomSec}

This section elaborates on the shared geometrical representation between OpenMC and DEGAS2: volumes bound by quadratic surfaces

The computational domain is divided into volumes of piecewise-constant properties known as \emph{cells}. Cells are practially differentiated by either different tallies, different background properties against which the flights interact, or both. Flights are advanced through these volumes step-wise: the straight-line trajectories are interrupted only by cell boundaries or by a scattering event. In the absence of scattering, the flight is advanced discretely to the next cell boundary. To do so efficiently, the cells are restricted to be bound by \emph{quadratic surfaces}. In general, a quadratic surface is defined by the equation:
\begin{equation}
   \sum\limits_{i=0}^{2} \sum\limits_{j=0}^{2-i} \sum\limits_{k=0}^{2-j} C_{ijk} x^i y^j z^k = 0,
\end{equation}
where the summation limits the equation to at most quadratic in the Cartesian coordinates $(x,y,z)$. A surface is therefore defined by the 10 coefficients $C_{ijk}$, and includes planes, cones, spheres, cylinders, paraboloids, and hyperboloids. The overall sign of the coefficients gives the surfaces' directionality (inside/outside, above/below, etc.) that serve to define the cells that they bound. However, this is not sufficient to uniquely determine a cell; often the surfaces intersect elsewhere in space. To disambiguate the cell, \emph{cut surfaces} are employed that have the same properties as cell faces but do not act as the cell's boundary. 

\begin{figure}
   \begin{center}
   \includegraphics[width=1.0\textwidth]{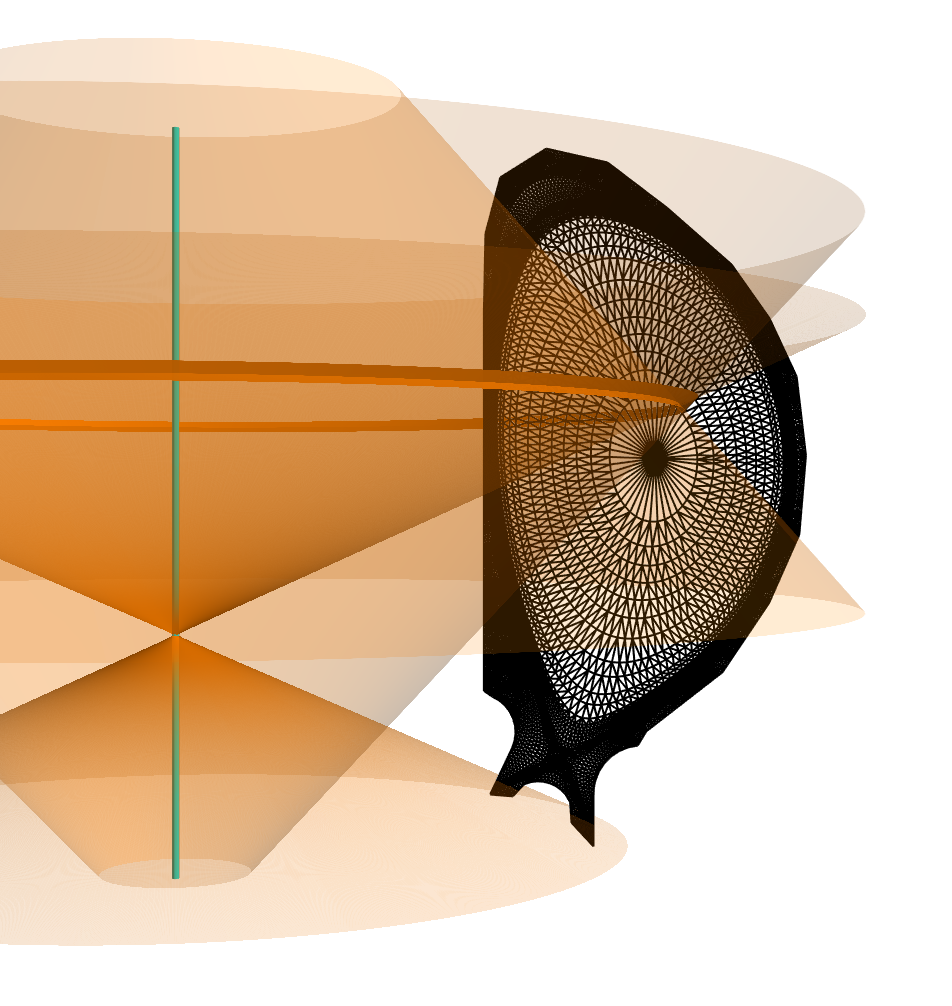}
   \end{center}
   \caption{\label{geomfig} An illustration of how volumetric cells are constructed in OpenMC and DEGAS2. Shown here is an ITER-like mesh partially aligned with flux surfaces (black lines). Transparent orange cones are quadratic surfaces: the basic geometrical element. The solid-orange ring is the volume thereby constructed by the intersection (when further constrained vertically). The vertical cyan line is the axis of rotation.}
\end{figure}

Consider, as we do in \cref{fullmeshSec}, the special case of a 2D triangular mesh rotated around an axis of symmetry (the $z$ axis), as illustrated in \cref{geomfig}. Each line segment that bounds a triangular element is a portion of a quadratic surface forming the face of the 3D cell. Most of these surfaces are cones with vertex along the $z$ axis, but special cases are cylinders and planes for vertical or horizontal segments, respectively. The equations specifying such surfaces are:
\begin{itemize}
   \item \textbf{Horizontal plane} at $Z_0$: $z - Z_0 = 0$
   \item \textbf{Cylinder} centered on $z$ axis with radius $R_0$: $x^2 + y^2 - R_0^2 = 0$
   \item \textbf{Cone} with a slope of $m$ and vertex at $z=Z_0$: $m^2 x^2 + m^2 y^2 - z^2 + 2 Z_0 z -Z_0^2 = 0$.
\end{itemize}
Each of these surfaces (cell faces), together with a sign, define a half space. Each cell in the domain is formed by the intersection of three such half spaces, plus additional ``cut surfaces'' to disambiguate the volume of intersection. 

With the computational geometry so specified, solving for the intersection of a straight-line trajectory with a surface face is computationally trivial. In the absence of scattering (see \cref{reactionSec}), when a flight begins at a coordinate on a cell's face its velocity is extended along a straight-line trajectory, and the distance to the intersection of each of the cell's bounding surfaces is calculated. The actual distance traversed is chosen as the smallest of these intersection distances, and the next position in the flight is calculated on the exiting surface.

\subsection{Reactions} \label{reactionSec}

The cases considered here involve two types of reactions: ionization and charge exchange. The former is analogous to neutron absorption and will be treated similarly. The latter is similar to elastic scattering, where the post-collision velocity is drawn directly from the target distribution.

Firstly, it is necessary to clarify what is meant by ``cross section'' in the context of atomic physics. The cross section $\sigma$ is a function of the particular test and target particles' relative velocity. Quantities averaged over the target velocity distribution are indeed relevant, but these are not generally referred to as cross sections. In contrast, the neutronics community and OpenMC in particular, use the term ``cross section'' to refer to population-averaged quantities and are thereby functions of target population temperature. To avoid the confusion that would accompany multiple definitions of the term, this work follows the former convention: the cross section is defined as that before velocity averaging is performed. 

The \emph{reaction rate} is denoted by:
\begin{equation}
   \left\langle \sigma_{ss'} v \right\rangle = \frac{1}{n_{s'}} \int u \sigma_{ss'}(u) f_{s'}\left( \mathbf{w} \right) \,\mathrm{d}^3\mathbf{w},
\end{equation}
which has units of $\mathrm{m}^3/s$ such that, upon multiplying by the target density $n_{s'}$, the collision frequency $\nu_{ss'} = n_{s'} \left\langle \sigma v \right\rangle$ is recovered. In OpenMC, the tabulated ``cross section'' is identified with $\left\langle \sigma_{ss'} v \right\rangle / v$, where $v$ is the test particle speed. 

Due to the large separation of scales for electron velocities relative to recycling atoms, DEGAS2 precalculates the reaction rate as a function of electron density and temperature, and is independent of atomic velocity. In this way, the cross section of electron-impact ionization is not used directly. In OpenMC, there is currently no way to specify a reaction rate that is exactly constant; it is calculated directly from the cross section. In order to mimic DEGAS2's treatment of ionization in OpenMC, we employ the ``Maxwell molecule'' form of the cross section such that $\sigma_{iz}(v) = \left\langle \sigma_{iz} v \right\rangle/v$. With the electron speed $w \approx u \gg v$, the integration over electron velocities decouples from $f_s(\mathbf{v})$ and we can write the collision operator as a loss term:
\begin{equation}
   C_{iz}\left[f_s, f_e \right] = - f_s \left(\mathbf{v}\right) \int w \sigma_{iz}\left(w\right) f_e\left( \mathbf{w} \right) \,\mathrm{d}^3 \mathbf{w} = - n_e \left\langle \sigma_{iz} v \right\rangle f_s\left( \mathbf{v} \right).
\end{equation}
For this case, $\left\langle \sigma_{iz} v \right\rangle$ is independent of $\mathbf{v}$ and is tabulated as a function of electron density and temperature in DEGAS2. In OpenMC, we treat the reaction as a ``radiative absorption'' (MT type 102 in the ENDF database \cite{brown2018endf}), with a tabulated cross section inversely proportional to the \emph{test particle} speed, yielding a constant reaction rate. With the atomic database in DEGAS2, this can be generalized such that the ``effective density'' of the electrons for ionization is proportional to the ionization reaction rate. In practice, the same result is obtained in OpenMC by treating $u=v$ (the test particle velocity) instead. As in the physical interpretation, the Maxwell molecule form of the cross section enables the relative velocity to be ignored, and thereby the integration over $\mathbf{w}$ is decoupled from $\mathbf{v}$.

For charge exchange, a new reaction type is needed. It is similar to elastic scattering in a material, except the final velocity is \emph{exchanged} with the velocity of the target rather than an elastic momentum exchange with a nuclide of non-zero velocity. The reaction dataset is otherwise prepared analogously to elastic scattering in OpenMC: the ``0K cross section'' directly corresponds to the atomic (DEGAS2) definition, while data at other ion temperatures is scaled by a factor proportional to the precalculated reaction rate. Such a reaction has been implemented in the fork of OpenMC used for the benchmarks that follow.

%Supplementary material includes the digital notebooks that generated the atomic data tables in OpenMC.

\section{OpenMC-DEGAS2 benchmarks} \label{sec:results}

This section demonstrates that, with relatively minor and unobtrusive adjustments to the code base, OpenMC can be used for atomic-plasma interaction while obtaining results that generally compare favorably with DEGAS2. 
%For these benchmarks, the test particle mass was adjusted slightly to match the mass of a hydrogen atom rather than a neutron. 

Performance was found to depend on the complexity of the domain, with some cases running faster than DEGAS2 and others running slower. Since the reaction table lookup in DEGAS2 is uniformly spaced either linearly or logarithmically, this is expected to be more efficient than in OpenMC, where the tables have more arbitrary structure (and thereby, less efficient interpolation). Fortunately, this does not appear to significantly affect the performance in OpenMC.

These benchmarks were run on a Lenovo T480 laptop with 4 dual-threaded cores. For these benchmarks, OpenMC uses OpenMP parallelization (with 8 threads in this case), and DEGAS2 uses MPI parallelization with 4 processes. Both codes make nearly full utilization of the CPUs after their respective initialization procedures. 

\subsection{Ionizing box} \label{izonlySec}

The simplest benchmark in this work simulates a 2 m cube with planar symmetry and specularly reflective boundaries. Along the $z$ axis, the domain is subdivided into 20 equal volumes on which the tallies are calculated. Hydrogen atoms are produced uniformly within the first sub-volume and ionize throughout the domain at a constant rate. The rate of ionization corresponds to an electron density of $10^{19}\, \mathrm{m}^{-3}$ and a temperature of 10 eV. An equivalent reaction for electron-impact ionization is prepared for OpenMC as described in \cref{reactionSec}. 

\begin{figure}
   \begin{center}
   \includegraphics[width=1.0\textwidth]{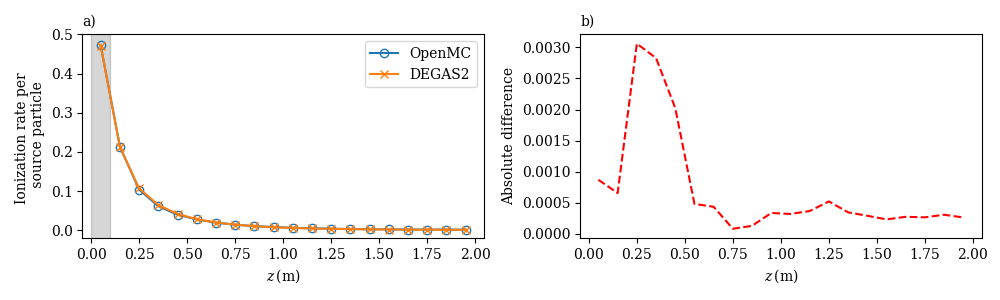}
   \end{center}
   \caption{\label{1dcaseFig}Example solution for uniform box case with only ionization with $N_p=10^4$ as solved by both DEGAS2 and OpenMC (a), with the absolute difference between the calculated solutions (b). The gray rectangle in (a) illustrates the first cell wherein particles are produced.}
\end{figure}

\begin{figure}
\begin{minipage}[c]{0.45\textwidth}
   \begin{center}
   \includegraphics[width=0.9\textwidth]{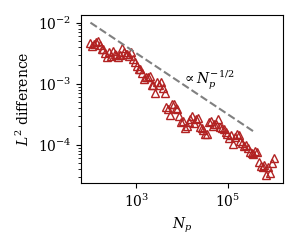}
   \end{center}
   \caption{\label{error-iz}Measure of the differences between the OpenMC and DEGAS2 solutions of the uniform box case with ionization only, shown for various numbers of simulation particles (flights) $N_p$ as an $L^2$ norm: $E = \sqrt{ \sum_{i=0}^{N_z} \left( \hat{R}_i^\mathrm{(OMC)} - \hat{R}_i^\mathrm{(D2)} \right)^2 } / N_z$.}
\end{minipage} \hfill
\begin{minipage}[c]{0.45\textwidth}
%   \end{figure}
%\begin{figure}
   \begin{center}
   \includegraphics[width=0.9\textwidth]{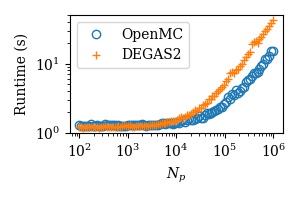}
   \end{center}
   \caption{\label{performance-iz}Time of execution for a series of Monte Carlo simulations of the uniform box case with ionization only for various numbers of particles/flights $N_p$. Not included in the counted time are geometry preparation or tally reading.}
\end{minipage}
\end{figure}

An example of the results for the ionization rate in each sub-volume is shown in \cref{1dcaseFig} for the case of 10,000 flights. Excellent agreement is seen, converging further as the number of flights increases. This is repeated for many different numbers of flights/particles $N_p$. \Cref{error-iz} shows the discrete $L^2$ relative difference between the two solutions, scaling as one would expect with $1/\sqrt{N_p}$. 

A comparison of the performance of OpenMC relative to DEGAS2 is displayed in \cref{performance-iz}, showing the execution time for each of the cases with different $N_p$. The main Monte Carlo calculation in OpenMC is faster than DEGAS2 by a factor of 2--3$\times$, except for when the number of flights is relatively low. 

\subsection{Box with charge-exchange}

\begin{figure}
   \begin{center}
   \includegraphics[width=1.0\textwidth]{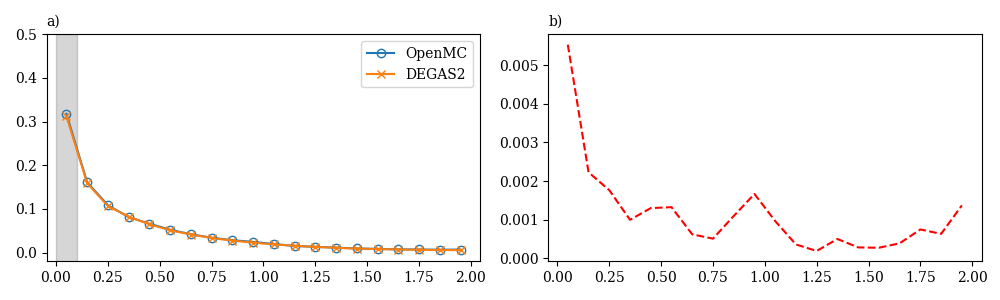}
   \end{center}
   \caption{\label{1dcaseFig-cx}Example solution for uniform box case with ionization and charge exchange with $N_p=10^4$ as solved by both DEGAS2 and OpenMC (a), with the absolute difference between the calculated solutions (b). The gray rectangle in (a) illustrates the first cell wherein particles are produced.}
\end{figure}

To isolate potential performance complications introduced by a reaction with nontrivial dependence on relative speed, the benchmark of \cref{izonlySec} is repeated with an additional reaction. In this section, an additional species of protons is introduced with which atoms interact via charge exchange. The protons have the same density as electrons ($10^{19}\,\mathrm{m}^{-3}$) and a temperature of 100 eV. This solution departs significantly from the ionization-only case, with the atoms penetrating deeper into the domain from the source region (see \cref{1dcaseFig-cx} for an example solution, again with $N_p = 10^4$).

\begin{figure}
\begin{minipage}[c]{0.45\textwidth}
   \begin{center}
   \includegraphics[width=0.9\textwidth]{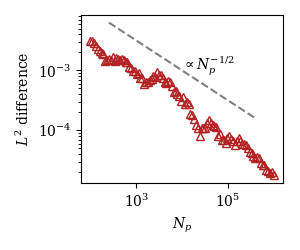}
   \end{center}
   \caption{\label{error-cx}Measure of the differences between the OpenMC and DEGAS2 solutions of the uniform box case with ionization and charge exchange. Shown for various numbers of simulation particles (flights) $N_p$ as an $L^2$ norm: $E = \sqrt{ \sum_{i=0}^{N_z} \left( \hat{R}_i^\mathrm{(OMC)} - \hat{R}_i^\mathrm{(D2)} \right)^2 } / N_z$.}
\end{minipage} \hfill
\begin{minipage}[c]{0.45\textwidth}
%   \end{figure}
%\begin{figure}
   \begin{center}
   \includegraphics[width=0.9\textwidth]{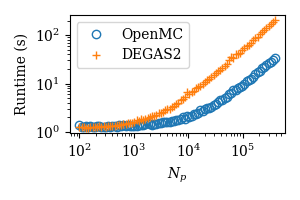}
   \end{center}
   \caption{\label{performance-cx}Time of execution for a series of Monte Carlo simulations of the uniform box case with ionization and charge exchange for various numbers of particles/flights $N_p$. Not included in the counted time are geometry preparation or tally reading.}
\end{minipage}
\end{figure}

Again, excellent agreement is seen between OpenMC and DEGAS2, converging even more closely than the ionization-only case (see \cref{error-cx}). The performance comparison (\cref{performance-cx}) is also even better, with OpenMC outperforming DEGAS2 by nearly an order of magnitude for large numbers of flights. This performance benchmark effectively puts to rest questions over whether OpenMC would reach adequate performance due to its more general spacing of cross section lookup tables.

\subsection{Realistic axisymmetric mesh} \label{fullmeshSec}

The success of the simple benchmarks motivates further investigation. We find that the improved performance of OpenMC does not translate directly to a case with less trivial discretized geometry without further optimization, though it remains comparable to DEGAS2. Excellent agreement between the two codes is again maintained.

\begin{figure}
   \begin{center}
   \includegraphics[width=0.6\textwidth]{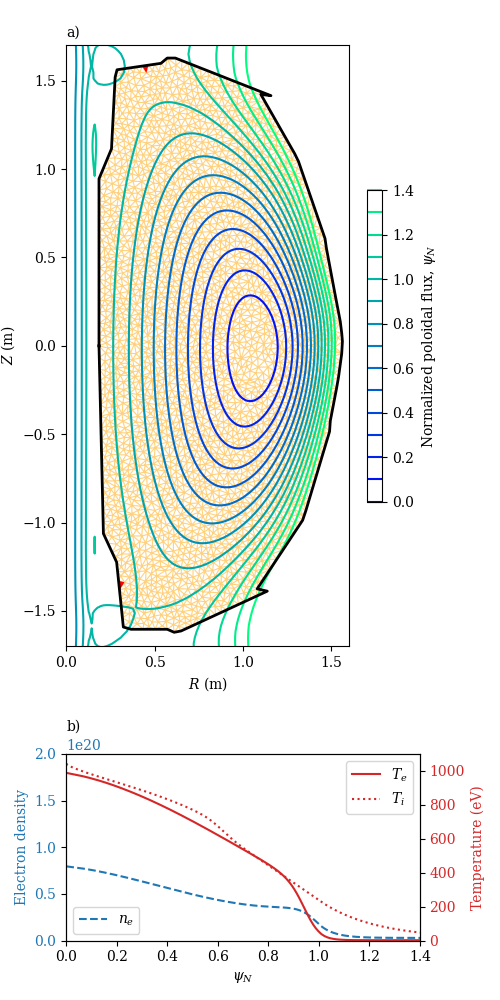}
   \end{center}
   \caption{\label{nstx-mesh} a) Illustration of the domain used for the NSTX-U benchmark. The black polygon represents the solid wall, orange triangles show the individual volumetric elements therewithin. Contours of constant $\psi_N$ are also shown. The two red triangles adjacent to the wall near the top and bottom are the volumes in which source particles are generated. b) shows the 1D plasma profiles of electron density (blue dashed), electron temperature (solid red) and ion temperature (dotted red) that get mapped onto the $(R,Z)$ plane via $\psi_N$.}
\end{figure}

This benchmark takes a more realistic discretization of a tokamak geometry, wherein a limiter shape similar to that of Upgraded National Spherical Torus Experiment (NSTX-U) in the $(R,Z)$ plane is triangulated with DEGAS2's utility (resulting in about 5,000 spatial elements). These triangular elements are rotated about the $z$ axis to form 3D cells bound by quadratic surfaces (as described in \cref{geomSec} and illustrated in \cref{geomfig}).  This same volumetric discretization is used for the OpenMC simulations. See \cref{nstx-mesh} for an illustration of this mesh.

The example plasma background is also common to the two codes in the benchmark, and is described as follows. Each triangular element is associated with its centroid's value of $\psi_N$, the normalized poloidal magnetic flux on which plasma properties---electron density $n_e(\psi_N)$, electron temperature $T_e(\psi_N)$ and ion temperature $T_i(\psi_N)$---are approximately constant. Contours of $\psi_N$ are shown in \cref{nstx-mesh}a. The first magnetic separatrix, dividing open and closed field lines in the domain, is characterized as $\psi_N=1$, while $\psi_N=0$ corresponds to the magnetic axis. The mapping of $(R,Z) \rightarrow \psi_N$ is a function of the magnetic geometry, and we take the reconstructed double-null equilibrium for NSTX-U discharge 205020 studied in Ref. \cite{scotti2018divertor}.

The plasma background was also calculated self-consistently with the magnetic equilibrium and appropriate diagnostics in Ref. \cite{scotti2018divertor}, with profiles shown in \cref{nstx-mesh}b. Although these NSTX-U discharges used deuterium, for consistency with the 1D benchmark and associated reactions, we take the plasma ions to instead be protons with a density $n_i = n_e$ such that there are no plasma impurities. 

Reactions included are electron-impact ionization and charge exchange, as described in \cref{reactionSec}. Neutral hydrogen atoms are produced with a Maxwellian source distribution at 3~eV within the two elements highlighted in red in \cref{nstx-mesh}a, corresponding approximately to the ``strike points'' where the plasma exhaust intersects the wall. The number of flights launched in each of these volumes is one million ($N_p = 2\times 10^6$ flights total). 

\begin{figure}
   \includegraphics[width=0.8\textwidth]{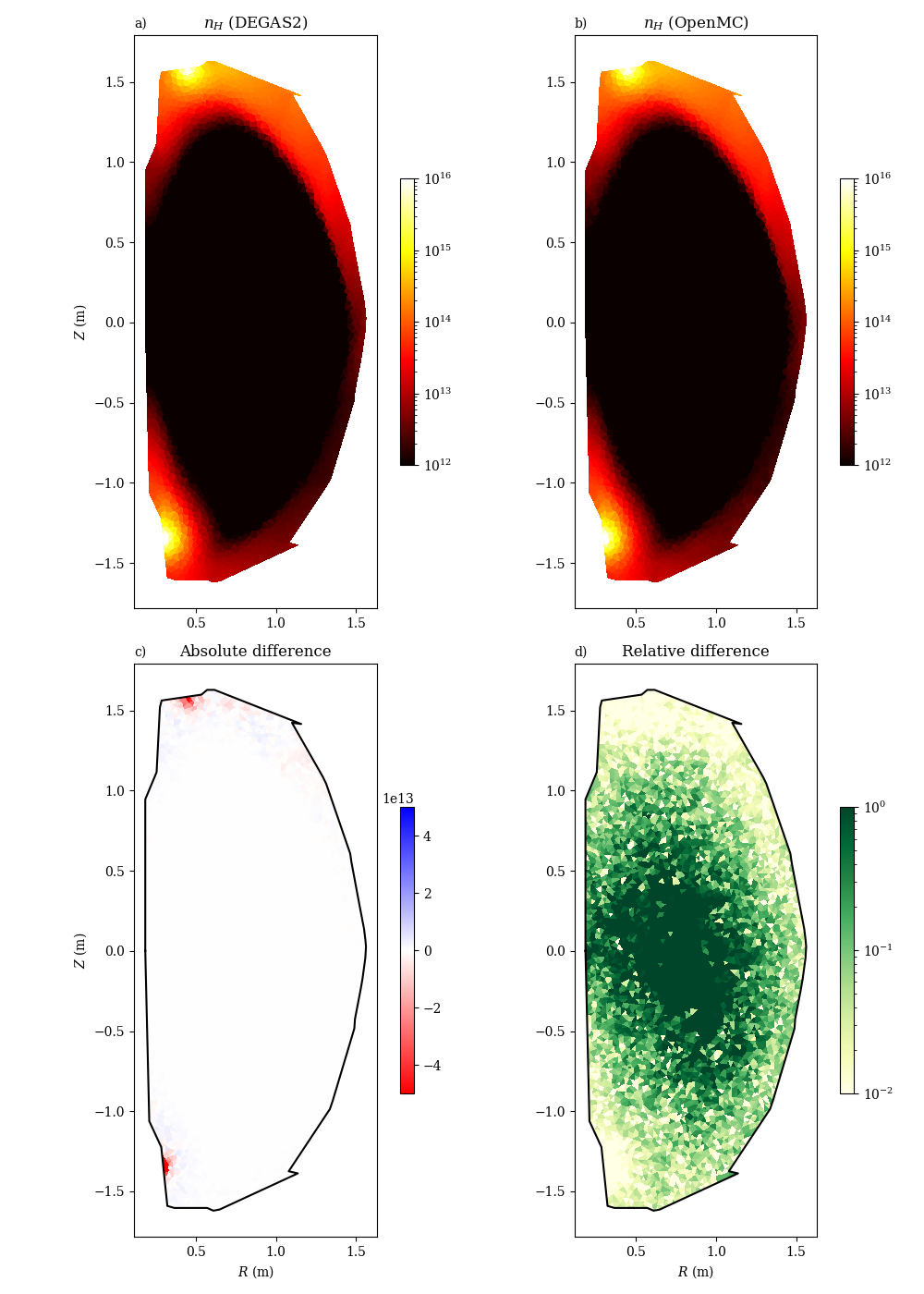}
   \caption{\label{2dresults}Comparison between calculated results with DEGAS2 and OpenMC. a) and b) show the direct results for atomic density in DEGAS2 and OpenMC, respectively. c) shows the absolute difference between the two, with d) the relative difference.}
\end{figure}

The predicted atomic density is shown in \cref{2dresults}, along with the absolute and relative differences between them. Again, excellent agreement between DEGAS2 and OpenMC is observed. The relative error only approaches unity in the core region where the neutral density (and absolute error) is low. As is typical, neutral atoms are mostly confined the relatively cool edge, where the lower electron density and temperature results in a lower ionization rate.

The performance of OpenMC in this benchmark is about 30\% slower than DEGAS2 (1308~s runtime versus 970~s, respectively). This is attributed to the large number of unstructured cells, each of which contains their own background reaction properties (effectively over 5000 unique ``materials'' in OpenMC parlance). It is anticipated that this performance shortfall can be mitigated by GPU acceleration and/or relatively straightforward optimizations.

Other similar cases were run with different background plasma and source volume locations, but are not shown here. These produced similar agreement and performance comparison; \cref{nstx-mesh,2dresults} are representative of these iterations.

\section{Future work} \label{sec:future}

The goal of this work was to illustrate that with modest changes OpenMC can provide accurate predictions of atomic transport under reactions with a plasma background while obtaining competitive performance. This is intended to motivate reproducing the functionality of established tools such as DEGAS2 and EIRENE and to test whether such a generalization is worthwhile. The success of our benchmarks indicate that this is indeed the case, since OpenMC would bring with it advanced capability such as GPU acceleration, CAD integration, modern software architecture, and an active open-source community. This will allow, for example, atomic simulations to be routinely included in digital twin models of reactors, as is currently being done with OpenMC for neutrons.

The tasks required to reproduce the capability of DEGAS2 follow. Firstly, OpenMC was designed strictly as a neutron transport solver and was later generalized to include photons. Becoming a feasible replacement for DEGAS2 requires a more flexible range of test particle species to include much of the periodic table. The associated reactions will also need to be generalized and read from existing databases. In this work, we used the existing OpenMC framework to reproduce the physics of atomic interactions. Ideally, multiple species would be present in a single simulation since some reactions transform one neutral particle species into another (e.g., excitation to a metastable state or dissociation of molecules). Other generalizations that are somewhat less cumbersome but nevertheless necessary are: a broader menu of tally options, a surface source option with a partially tabulated velocity distribution (to model material desorption and prompt reflection together), synthetic line radiation diagnostics, and iterative nonlinear intra-neutral scattering.

We have demonstrated that, with these upgrades, atomic and molecular modeling, especially in the context of nuclear fusion, have a clear path to sustainability and high performance, while maintaining the accuracy, robustness, and, eventually, the generality of existing legacy tools.

\section*{Acknowledgments}

The authors are grateful to J. Menard for supporting this work, and to J.-M. Kwon and R. Akers for motivating discussions. 
This work was supported by the US Department of Energy Office of Science under award number  DE-AC02-09CH11466. 
%Disclaimer: This report was prepared as an account of work sponsored by an agency of the United States Government. Neither the United States Government nor any agency thereof, nor any of their employees, makes any warranty, express or implied, or assumes any legal liability or responsibility for the accuracy, completeness, or usefulness of any information, apparatus, product, or process disclosed, or represents that its use would not infringe privately owned rights. Reference herein to any specific commercial product, process, or service by trade name, trademark, manufacturer, or otherwise does not necessarily constitute or imply its endorsement, recommendation, or favoring by the United States Government or any agency thereof. The views and opinions of authors expressed herein do not necessarily state or reflect those of the United States Government or any agency thereof. 

\bibliographystyle{unsrt}
\bibliography{zotero}

\end{document}